\documentstyle[12pt]{article}
\renewcommand{\baselinestretch}{0.95}

\title{\vspace*{-1cm}
{\Large \bf Wiener Integration for Quantum Systems:\\
A \hspace*{-0.4mm}Unified Approach to the Feynman-Kac formula}
\thanks{Partially based on a plenary lecture given
by H. L. at the international conference on path integrals,\/\
Dubna, Russia, May 27--31, 1996; in: {\it Path integrals: Dubna '96}, 
eds. V. S. Yarunin and M. A. Smondyrev, JINR E96--321, ISBN: 5--85165--451--1,
Dubna
1996, pp. 95--106.}}
\author{\large Bernhard Bodmann,
Hajo Leschke and Simone Warzel \\[3mm]
\normalsize\em Institut f\"ur Theoretische Physik\\
\normalsize\em Universit\"at Erlangen-N\"urnberg\\
\normalsize\em Staudtstra\ss e\ 7, D-91058 Erlangen, Germany}
\date{}

\newcommand\ba{\begin{array}}
\newcommand\ea{\end{array}}
\newcommand\be{\begin{enumerate}}  
\newcommand\ee{\end{enumerate}}  
\newcommand\bi{\begin{itemize}}  
\newcommand\ei{\end{itemize}}  
\newcommand\bd{\begin{description}}  
\newcommand\ed{\end{description}}  
\newcommand\beq{\vspace*{-1mm}\begin{equation}}  
\newcommand\eeq{\end{equation}}  
\newcommand\beqa{\vspace*{-2mm}\begin{eqnarray}\vspace*{-2mm}}  
\newcommand\eeqa{\end{eqnarray}}  

\newcommand\libitem[2]{\bibitem{#1}  #2}

\newcommand{\eq}[1]{(\ref{#1})}

\newcommand\eqbox[1]{\beq\fbox{\rule[-.8em]{0em}{2.3em}$\displaystyle\ #1\ $}
\eeq}
\newcommand\DS{\displaystyle}
\newcommand\TS{\textstyle}

\renewcommand\={\!\!=\!\!}

\newcommand\one{{\hat 1}}                 
\newcommand\A{{\op A}}
\newcommand\B{{\op B}}

\newcommand\F{{\op F}}

\renewcommand\H{{\op H}}                       
\renewcommand\P{{\hat p}}                        
\renewcommand\S{{\op S}}                       
\newcommand\Q{{\hat q}}
\newcommand\R{{\op R}}
\newcommand\U{{\op U}}
\newcommand\p{{\hat p}}            
\newcommand\q{{\hat q}}            
\newcommand\T{{\op T}}

\def\rta{\rightarrow}
\def\absq#1{\left| #1 \right|^2}
\def\ifmath#1{\relax\ifmmode #1\else $#1$\fi}
\def\half{\ifmath{{\textstyle{1 \over 2}}}}

\newcommand\abs[1]{\vert {#1}\vert}
\newcommand\av[1]{\langle#1\rangle}
\newcommand\bra[1]{\langle#1|\,}

\newcommand\dby[1]{{d\over d#1}}
\newcommand\delby[2]{{\partial #1 \over \partial #2}}

\newcommand\ket[1]{\,|#1\rangle}

\newcommand\op[1]{\widehat{#1}}

\renewcommand\a{\alpha}                          

\newcommand\de{\delta}

\newcommand\n{\nu}

\renewcommand\dot[1]{\stackrel{\mkern2mu\mbox{\bf .}}{\raisebox{-0pt}{$#1$}}
\mkern-5mu}

 \newlength{\nn}\newlength{\ii}
 
 \newlength{\zz}
 
 \newlength{\rr}
 \def\rz{\,\mbox{$\mbox{R}\hspace{-1.23\rr}\mbox{I}\hspace{-1\ii}
 \hspace{1.23\rr}$}}
 \newlength{\rrs}
 \newlength{\iis}
 \def\krz{\mbox{$\hspace{0.3\rrs}\mbox{{\scriptsize R}}\hspace{-1.25\rrs}%
  \mbox{{\scriptsize I}}\hspace{-1\iis}%
  \hspace{1.25\rrs}$}}
 \newlength{\cc}\newlength{\iii}%
 \def\str{\mbox{$\scriptstyle |$}}
 \def\cz{\mbox{$\mbox{C}\hspace{-.7\cc}\mbox{\raisebox{0.45ex}{\str}}%
 \hspace{-1\iii} \hspace{.7\cc}$}}%

\newcommand\Bbb[1]{\rz}

\def\ibid{{\it ibid.\/}}

\def\@eqnsel\hfil

\newskip\@eqnmar
\@eqnmar\@centering

\let\@eqnhook\relax

\def\eqnarray{\stepcounter{equation}\let\@currentlabel=\theequation
\global\@eqnswtrue
\global\@eqcnt\z@\tabskip\@eqnmar\let\\=\@eqncr
\@eqnhook
$$\halign to \displaywidth\bgroup\@eqnsel
  \vspace*{-2mm}$\displaystyle\tabskip\z@{##}$&\global\@eqcnt\@ne
  \hfil${{}##{}}$\hfil&\global\@eqcnt\tw@
  \vspace*{-2mm}$\displaystyle{##}$\hfil
  \tabskip\@centering&\llap{##}\tabskip\z@\cr\vspace*{-0.5mm}}

\newcommand\FK{Feyn\-man--Kac}

\newcommand\Dw{{Dw}}
\newcommand\IDw{{\int\!\Dw\,}}

\newcommand\ep[1]{{\rm e}^{\,\mbox{\xpt $ #1 $}}}

\newcommand\Tt{{\T_t}}
\newcommand\Ut{{\U_t}}
\newcommand\Ttw{{\Tt\left(w\right)}}
\newcommand\Utw{{\Ut\left(w\right)}}
\newcommand\Usw{{\U_s\left(w\right)}}

\newcommand\dw{{ds \dot w(s) }}
\newcommand\texp[1]{\stackrel{\TS \longleftarrow}{\exp}\left\{  #1 \right\} }
\newcommand\Ids[1]{\int_0^{#1} \! ds \,}
\newcommand\Idsn[2]{\int_0^{#1} \! ds_{#2} \,}
\newcommand\Idw[1]{\int_0^{#1} \! \dw}

\newcommand\oa{ \bigcirc \!\!\!\!\! \alpha \!}

\begin{document}

\maketitle

\begin{abstract}
A generalized Feynman--Kac formula based on the Wiener measure is presented. 
Within the setting of a quantum 
particle in an electromagnetic field it yields the standard Feynman--Kac 
formula for the
corresponding Schr\"odinger semigroup. In this case rigorous criteria for 
its validity are compiled. Finally, 
phase--space path--integral representations for more general quantum 
Hamiltonians are derived.
These representations rely on a generalized Lie--Trotter formula which 
takes care of the operator--ordering 
multiplicity, but in general is not related to a path measure. 
\end{abstract}

\vspace{2mm}
\hspace*{\fill}\begin{minipage}[t]{12cm}%
{\it Actually, in the Wiener integral the things are much simpler.}\ 
\cite{Zel90}
\end{minipage}


\section{The Feynman--Kac formula, revisited}
 
More than seventy years of Wiener's path integration have been most 
gratifying for both mathe\-maticians and
theoretical physicists. Originally constructed as a mathematical model 
for the phenomenon of
Brownian motion, it nowadays plays a major r\^ole also in polymer and 
quantum physics, and still is
fundamental to the theory of general stochastic processes \cite{ReYo94}. 
The importance of Wiener's measure  \cite{Wie23} for quantum physics 
became clear soon after Feynman's stimulating
paper \cite{Fey48} when Kac \cite{Kac49, Kac} identified it as the 
key to a probabilistic representation of 
Schr\"odinger semigroups, these days called the Feynman--Kac formula 
\cite{Sim79,GlJa87,Roe94}.

In this section we intend to survey essentials of this formula and 
some of its relatives. In doing so,
we leave aside most  technicalities and subtleties which we believe 
to be of secondary
importance for applications, in particular when they tend to obscure 
the simplicity of the underlying
ideas.\pagebreak

\subsection{The Wiener measure and stochastic integrals}

The standard {\it Wiener measure}
$\Dw$ is a certain probability distribution on the space of 
continuous paths $\{w \! : \rz_+ \rightarrow \rz^d, s \mapsto w(s)\}$ from the
positive half--line \mbox{$\rz_+:=[0,\infty[$} into
$d$--dimensional Euclidean space $\rz^d$ which start at the origin, 
$w(0)=0$. It is uniquely determined and
well--defined by the following properties:
\begin{itemize}
\vspace*{-2.5mm}
  \item[i)] 
                $\Dw$ is a Gaussian probability measure.
\vspace*{-2.5mm}
\item[ii)] 
    Its first and second moments are
           \vspace*{-0.5mm}
           \beq
                        \IDw w_j(s) =  0 \, , \qquad
                        \IDw w_j(r) w_k(s)  =  \de_{jk} \min(r,s) \,. 
\label{gvar}
           \eeq
           \vspace*{-0.5mm}
\end{itemize}
\vspace*{-6mm}
Here and in the following indices $j,k,\dots$ denumerate the components 
of $d$--com\-po\-nent quantities,
$x\cdot y := \sum_{j=1}^d x_j y_j $ denotes the standard scalar product 
of two such quantities, and the usual convention
$x^2:=x \cdot x$ is adopted. Kronecker's delta in \eq{gvar} 
expresses the stochastic independence of different components of $w$. A
one--line characterization of $\Dw$, equivalent to i) and ii),  
is given by its (functional) Fourier transform
\eqbox{ \label{Wiener} \mbox{\rule[-1em]{0em}{2.6em}}
 \IDw \exp\biggl\{-i \int_0^\infty \! \! ds \, w(s) \! \cdot \! f(s) 
\biggr\} = 
      \exp\biggl\{-\half \int_0^\infty \! \! dr \! \int_0^\infty \! \! ds 
\, \min(r,s) f(r) \! \cdot \! f(s) \biggr\}   }
where $f \! : \rz_+ \rightarrow \rz^d$ is any $d$--component function 
with compact support in the half--line $\rz_+$.

Whenever the Wiener measure $\Dw = \prod_{1\leq j \leq d} \Dw_j$ is 
applicable, its advantages over lattice or time--slicing
prescriptions for path integration can hardly be overestimated. 
Since $\Dw$ is a positive, countably additive and normalized measure,
the powerful machinery of general integration and probability theory 
\cite{Lan93,Bau96} is at hand for the computation or estimation
of Wiener integrals $\int \! \Dw\, F(w)$, where $F$ is a functional 
of the paths. The importance of $\Dw$ for integration 
in infinite--dimensional path space $\bigl(\rz^d\bigr)^{\raisebox{-2.5pt}
{$\krz_+$}}$ is similar to that of Lebesgue's measure $dx=\prod_{1\leq j 
\leq d} dx_j$
for `ordinary' integration in $\rz^d$.

For an intuitive understanding of $\Dw$ it is sometimes helpful 
to think of it as an
infinite--dimensional Lebesgue measure with Gaussian weight
\vspace*{-0.5mm}
\beq \label{nonsense}
\prod_{ 0<s \atop 1\leq j\leq d} \left( \frac{dw_j(s)}{\sqrt{2\pi 
\Delta s}} \exp\left\{-\frac{\Delta s}{2} \dot{w}_j(s)^2 \right\} 
\right) \vspace*{-0.5mm}
\eeq
in the limit $\Delta s \downarrow 0$. Strictly speaking,  
expression \eq{nonsense}
is meaningless because there is no translational invariant measure 
in infinite dimensions. In addition,
{\em Wiener paths} almost surely have no time derivative 
$\dot{w}\,\,\, := \dby{s} w$ for any $s$ and hence
are not of bounded variation on any compact interval $[0,t]$, 
see e.g.\ \cite{Bau96}.
Nevertheless, the concept of {\it stochastic integrals} allows 
to give precise meaning in a suitable probabilistic sense to 
(line) integrals of the form
\beq \label{stoint}
\int_0^t \! \dw \cdot g\left(w(s),s\right) \, , \quad t\geq 0
\eeq\pagebreak
for a wide class of functions $g:   \rz^d \times \rz_+ \rightarrow \rz^d$.
In principle, there is a huge multiplicity of rigorously 
definable candidates \cite[p.\ 136]{ReYo94} for
\eq{stoint}. As a rule however,
literature \cite{Arn74,McG74, ReYo94} discusses at most a
one--parameter subclass of stochastic integrals, denoted by us as
\vspace*{-0.5mm}
\beq
\oa \int_0^t \! dw(s) \cdot g\left(w(s),s\right) \, , \quad \a\in [0,1] \, . 
\eeq
While probabilists usually use $\a=0$ corresponding to It\^o's 
original proposal, we will 
here follow Stratonovich in choosing $\a=1/2$ for the interpretation 
of \eq{stoint}, so that the rules of ordinary calculus in the sense of 
Newton and Leibniz 
formally apply.
By the conversion
formula \cite[p.\ 113]{McG74} \cite[p.\ 136]{ReYo94}
\beq \label{conversion}
 \int_0^t \! \dw \! \cdot \! g\left(w(s),s\right) = 
\oa \int_0^t \! dw(s) \! \cdot \! g\left(w(s),s\right) + 
     \left(\half - \a \right) \int_0^t \! ds \left( \nabla \! 
\cdot \! g \right) \left(w(s),s\right) 
\eeq
the Stratonovich stochastic integral is related to the $\a$--stochastic 
integral.
Note that the second integral on the right--hand side containing 
the divergence of $g$ is understood in the Lebesgue sense
because of the continuity of Wiener paths. Stochastic integrals 
can also be defined for functions $(w,s) \mapsto g(w,s)$ which in 
their first argument may depend 
on the whole history
$\{w(r)\}_{r\le s}$ of the path up to the time value of the second 
argument \cite{Arn74, McG74, ReYo94}. Such non--anticipating functions 
might take values in $\rz^d$, in a Hilbert space or even in a
set of Hilbert--space operators \cite{GlJa87}.

As a simple application of the preceding discussion one obtains
\beq \label{whitenoise}
\IDw \exp\biggl\{-i \int_0^\infty \! \dw \cdot f(s) \biggr\} = 
\exp\biggl\{- \half  \int_0^\infty \! ds \, f(s)^2 \biggr\}
\eeq
for compactly supported $f$. This follows from (formal) integration 
by parts, equation \eq{Wiener} and the representation 
$\de(r-s)=\frac{\partial^2}{\partial r \partial s} \min(r,s)$ of Dirac's 
delta function. In analogy to \eq{Wiener},
equation \eq{whitenoise} may serve as a one--line characterization of 
$d$--component {\it{Gaussian white noise}}.

\subsection{A generalization of the Feynman--Kac formula}

The main purpose of this subsection is to extend equation \eq{whitenoise} 
to certain
operator--valued functions $f$.  Accordingly, let $\A = ( \A_1, \dots, \A_d ) 
$ be
a $d$--component operator acting on some separable Hilbert space $\cal H$. 
Moreover,
let $\B$ be another operator and $\one$ the identity on $\cal H$. 
For notational simplicity
we will assume that $\A$ and $\B$ are time--independent, 
the extension to the time--dependent case
being straightforward. Finally, let $\Ttw$ be the operator solving the 
linear {\em stochastic integral equation}
\beq \label{SIE}
  \Tt = \one - i \int_0^t \! \dw \! \cdot \! \A \, \T_s - \B \int_0^t \! ds 
\, \T_s 
\eeq
which is the precise formulation of the linear (Stratonovich) stochastic 
differential equation
\beq
  \dby{t} \Tt = - i  \left( \dot w(t) \cdot \A - i  \B \right) \Tt \, , 
\qquad \T_0 = \one \, .
\eeq\pagebreak
In analogy  to quantum dynamics with explicitly time--dependent 
Hamiltonians the 
solution $\Ttw$ can be obtained by iterating \eq{SIE} 
\beqa \label{Dseries}
  \Ttw  & \! = & \one + \sum_{n=1}^\infty (- i)^n \Idsn{t}{n} 
\left( \dot w(s_n) \cdot \A - i  \B \right) \Idsn{s_n}{n-1} \nonumber \\
           & & \qquad \qquad  \quad \times\left( \dot w(s_{n-1}) 
\cdot \A - i  \B \right)    \cdots \Idsn{s_2}{1}  \left( \dot w(s_1) 
\cdot \A - i  \B  \right) \\
           & \! =: \! & \texp{ - i \int_0^t \! ds \left( \dot w (s) 
\cdot \A - i \B \right) }  \vspace*{-1mm} 
\eeqa
which defines  Dyson's time--ordered exponential function 
$\stackrel{\TS\longleftarrow}{\exp}$, also known as product integral.

Now we are prepared to state a {\it generalized \FK\ formula} 
\eqbox{ 
              \label{FK}  \IDw \Ttw = \exp\left\{ -t \left( 
\half \A^2 + \B\right) \right\} \, . }
As we will see in the next subsection, the standard 
Feynman--Kac formula follows by a suitable choice of 
${\cal H}$, $\A$ and $\B$.
A heuristic argument for the validity of \eq{FK} goes as follows:
In the presence of time--ordering, $\B$ and all components of 
$\A$ may be treated as pairwise commuting \cite{Fey51,Kub62},
so that a formal application of \eq{whitenoise} to \eq{Dseries} 
with $ f(s) = {\mit \Theta}(t-s) \A $ gives the desired result.
Here ${\mit \Theta}$ stands for Heaviside's unit--step function.

A more convincing argument for \eq{FK} proceeds in two steps.
First consider the case of vanishing $\B$. Wiener integrating both 
sides of \eq{SIE}
and supposing
\beq \label{Nov}
  \Bigl\langle \Ids{t} \dot w_j(s) \T_s \Bigr\rangle = - 
\mbox{$\frac{i}{2}$} \A_j \Bigl\langle \Ids{t} \T_s \Bigr\rangle \, ,
\eeq
where for convenience the notation $ \av{\cdot} := \IDw ( \cdot ) $ is 
introduced, yields the integral equation
\vspace*{-0.5mm}
\beq
  \langle \Tt \rangle = \one - \half \A^2 \Ids{t}{} \langle \T_s \rangle
\eeq 
which, in turn, gives
$
  \langle \Tt \rangle = \exp\bigl\{ - \frac{t}{2} \A^2 \bigr\}   \,  .
$
Equation \eq{Nov} follows from 
integration by parts with respect to $\Dw$ using its Gaussian nature 
(cf.\ \cite[p.\ 32 and solution of exercise 1.8.22]{RyKrTa})
and the convention ${\mit \Theta}(0) := 1/2 $ corresponding to the 
Stratonovich interpretation.
For a rigorous proof of \eq{Nov} one should use Malliavin's stochastic 
variational calculus
\cite{Nua95, Mal}.
Clearly, if the components of $\A$ commute, time ordering can be dropped,
so that the result is obtained directly from $\Ttw = \exp\left( - i w(t) 
\cdot \A \right)$ and \eq{Wiener}.

The case $\B \neq \op 0$ will now be treated in the second step by a 
perturbative argument in the spirit of Kac's original proof for the 
standard Feynman--Kac formula (see \cite{Kac,Sim79,Roe94}).  For this 
purpose, we define
\beq
  \Utw := \,\, \texp{ - i \Ids{t}{} \dot w(s) \cdot \A }
\eeq\pagebreak
and
\beqa
  \S_t(w) & \! := &  \texp{ - \Ids{t} \left(\Usw\right)^{-1} \B \, \Usw } \\
              & \! = & \one - \Ids{t}  \left(\Usw\right)^{-1} \B \, \Usw \, 
\S_s(w) \,  .
\eeqa
Then, according to standard time--dependent perturbation theory
\beq \label{US}
  \Tt = \Ut\, \S_t = \Ut - \Ids{t} \Ut \, \U_s^{-1} \B \, \T_s \, .
\eeq
We now Wiener integrate both sides, assume the identity
\vspace*{0.5mm}
\beq \label{factorize}
 \langle \Ut \, \U_s^{-1} \B \, \T_s \rangle = \av{\U_{t-s}} \B \av{\T_s}
\eeq
and recall $ \av{\Ut} = \exp\left\{ - \frac{t}{2} \A^2 \right\} $ from the 
first step to obtain the Duhamel--type 
of integral equation
\beq \label{Duhamel}
     \av{\Tt} = \exp \left\{ - \mbox{$\frac{t}{2}$} \A^2 \right\}  - \Ids{t} 
\exp\left\{ - \mbox{$\frac{t-s}{2}$} \A^2 \right\} \B \av{\T_s} \, .
\eeq
This implies the initial--value problem
\beq
   \dby{t} \av{\Tt} = - \left( \half \A^2  + \B \right) \av{\Tt} \, , 
\qquad \av{\T_0} = \one
\eeq
which finally gives \eq{FK}. The claimed stochastic independence underlying 
\eq{factorize} is basically due to 
the Markov property of $\Dw$
and the fact that Wiener paths start continually afresh \cite{Sim79,Bau96}.

We conclude this subsection with four remarks:
\bi
\item Clearly, the Markov property of the Wiener measure $\Dw$ is 
responsible for the emergence of a {\em one--parameter
operator semigroup} on the right--hand side of the generalized Feynman--Kac 
formula \eq{FK}. The Gaussian nature of
$\Dw$ causes (the negative of) its  generator $\half\A^2 + \B$ to be 
quadratic in $\A$.
If $\A$  and $\B$ are Hermitian, the left--hand side manifestly is, 
because $\Dw$ does not change under reflection $ w(s) \mapsto -w(s) $.
\item Conversely, reading \eq{FK} from right to left may be viewed as 
uncompleting squares of
operators in an exponent
by {\it Gaussian linearization}. It is therefore closely related to the 
so--called Hubbard--Stratonovich trick, widely used
in different branches of theoretical physics. For applications 
to certain many--body systems
see the surveys \cite{Mue75,KoKuSm82}
and references therein. To our knowledge, so far mainly the case of 
commuting $\A$--components has been considered in this
context.
There is also a noteworthy similarity to Feynman's disentangling 
formalism \cite{Fey51,NaShSt}.
\item A simple corollary of \eq{FK} for  $d = 2$, namely 
\vspace*{-0.5mm}
\begin{eqnarray*}   
                                 \IDw \texp{ -i \Ids{t} \left[ 
\left(\dot w_1(s) + i \dot w_2(s)\right) \A_+  +
                                            \left(\dot w_1(s) - i 
\dot w_2(s)\right) \A_- - i \B\right] } 
\end{eqnarray*}
\parbox{8cm}{\vspace*{-4mm}$ \hspace*{0.5cm} \DS   =  \, \exp\left\{ - t 
\left( \A_+ \A_- + \A_- \A_+ + \B \right) \right\} \, , $} 
\hfill \parbox{2cm}{\vspace*{-4mm}\beq \label{product} \eeq}\newline\newpage
follows from the non--commutative extension
\vspace*{-0.5mm}
\beq
     \A_+ \A_- + \A_- \A_+ = \half \left( \A_+ + \A_-\right)^2 - \half 
\left( \A_+ - \A_-\right)^2 \vspace{-1.5mm}
\eeq
of the binomial formula for a pair  of operators $\A_+$ and $\A_-$.
This shows that one can also linearize symmetrized products of two, 
in general non--commuting operators.
\item In case of an infinite--dimensional Hilbert space $\cal H$ the 
arguments and computations in our derivation are
rather formal, unless one specifies a class of admissible operators 
$\A$ and $\B$ which is sufficiently wide to be of interest
for applications. In particular, domain questions require additional
efforts. See, for example, subsection 1.4 below.
\ei

\subsection{The standard \FK\ formula}

In this subsection we specialize to the usual setting  for 
non--relativistic quantum mechanics of an electrically
charged point mass in $\rz^d$ without internal degrees of freedom. 
Consequently, let 
$L^2(\rz^d)$ be the Hilbert space of complex--valued functions 
$\psi: \rz^d \rta \cz, q \mapsto \psi(q)$,
which are square--integrable with respect to the Lebesgue measure $dq$. 
Consider now a vector
potential
$a: \rz^d \rta \rz^d$ and a scalar potential $v: \rz^d \rta \rz$, both 
sufficiently well--behaved.
They give rise to a standard Hamiltonian, or Schr\"odinger operator, 
$ \H := \half \left(\p - a(\q)\right)^2 + v(\q) $ on $L^2(\rz^d)$
with magnetic field $ b_{jk} := \delby{a_j}{q_k} - \delby{a_k}{q_j} $.
Here $\p$ and $\q$ denote the usual self--adjoint momentum and position 
operators.
For notational simplicity, we choose units such that Planck's constant 
$\hbar$, the mass and
the charge of the particle are all unity.

Then the {\it standard Feynman--Kac formula} with a vector potential reads
\eqbox{ \label{FKS} \mbox{\rule[-1em]{0em}{2.6em}}
 \!\ep{\TS -t\H} = \IDw \ep{-i w(t) \! \cdot \! \p} \exp\left\{ i \Idw{t} 
\! \cdot \! a(\q + w(s)) - \Ids{t} v( \q + w(s))\right\}
    \!\!
} 

In fact, it is a special case of \eq{FK} for the choice 
${\cal H}=L^2(\rz^d)$, $\A = \p - a(\q)$ and $\B = v(\q)$. 
This follows from the first
equation in \eq{US} and the identity
\beqa
     \Utw & = & \texp{ - i \Idw{t} \cdot (\p - a(\q)) } \\
              & = & \ep{ - i w(t) \! \cdot \! \p } 
\exp\left\{  i \Idw{t} \cdot a(\q + w(s)) \right\}
\eeqa
which is again understood with the help of time--dependent perturbation 
theory in combination with the
fact that the time--dependent operator components
\beq
   \ep{i w(s) \! \cdot \! \p} \, a_j(\q) \, \ep{- i w(s) \! 
\cdot \! \p} = a_j(\q + w(s)) 
\eeq
commute for all values of $j$ and $s$. The specialization is completed 
by observing a similar shift--identity for
the scalar potential in $\S_t(w)$.

Several remarks apply: 
\bi
\item The Wiener integral has managed to completely disentangle 
the non--com\-mut\-ing operators $\p$ and $\q$ in 
the {\it Schr\"odinger semigroup} $\exp(-t\H)$.
The standard \FK\ formula in its basis--independent version \eq{FKS} 
has been presented before (cf. e.g. \cite{ GaSc89, AdGeLe84, BrHeLe89}).
Apart from  mathematical subtleties,
this abstract way of looking at the \FK\ formula extracts its essential 
ingredients and is often a convenient starting point for further
manipulations.
\item The Stratonovich stochastic integral in \eq{FKS} may be 
replaced by the corresponding $\a$-stochastic
integral using the conversion formula \eq{conversion}. The resulting 
formula might be called the {\it \FK--$\a$
formula}
extending a terminology introduced by Simon \cite{Sim79}. The correct 
unitary transformation $\ep{-t\H} \mapsto
\ep{i \chi(\q)} \ep{-t\H} \ep{-i \chi(\q)}$  under a change of gauge 
$a \mapsto a + \nabla \chi$ is most easily obtained from \eq{FKS} by 
using the rules
of ordinary calculus.
\item By applying both sides of \eq{FKS} to a wave function 
$\psi \in L^2(\rz^d)$ it is straightforward to
derive the usual probabilistic expression for its image $\exp(-t\H)\psi$ 
under the Schr\"odinger
semigroup (cf. \cite[eqs. (6.8) and (15.1)]{Sim79})
\vspace*{0.5mm}
\beqa \label{I}
  \Bigl(\ep{-t\H} \psi \Bigr)(q)  
 &  = & \IDw  \exp\left\{-i\Idw{t} \cdot a(q + w(s))\right\}  \nonumber \\
          &  & \quad \quad \times       \exp\left\{- \Ids{t} v(q + w(s))
\right\} \psi(q +w(t))  \, . \vspace*{-1mm}
\eeqa
This can be rewritten as
\beq
 \Bigl(\ep{-t\H} \psi \Bigr)(q)  = \int\! dq' \, \bra{q} \ep{-t\H} \ket{q'} 
\psi(q')
\eeq
where the Euclidean propagator, that is,  the position representation of the 
semigroup, is given by
\beqa \label{II}
\bra{q} \ep{-t\H} \ket{q'} &=&  \IDw \delta(w(t) + q - q') 
\exp\left\{-i\Idw{t} \cdot a(q + w(s))\right\} \nonumber \\
                                     &  & \quad \quad \times 
\exp\left\{-\Ids{t} v(q + w(s))\right\} \, .
\eeqa
Here, due to the presence of the delta function, effectively only 
those Wiener paths contribute which
pass through $q'-q$ at time $t$. One may also integrate over the 
rigidly shifted paths $s \mapsto 
w(s) + q$ which start at $q$ and pass through $q'$ at time $t$.

\item As a final point in this subsection, we mention a close relative of 
\eq{I}, the probabilistic
representation of the unitary time evolution
\vspace*{0.7mm}
\beqa \label{uni}
  \Bigl(\ep{-i t\H} \psi \Bigr)(q)  
 &  = & \IDw  \exp\left\{-i \sqrt{i} \Idw{t} \cdot a(q + \sqrt{i} \, w(s))
\right\}  \nonumber \\
          &  & \; \quad \times       \exp\left\{- i  \Ids{t} v(q + \sqrt{i} 
\, w(s))\right\} \psi(q +
\sqrt{i} \, w(t))  \, . \quad \vspace*{-0.5mm}
\eeqa\mbox{}\vspace{-0.5mm}%
It is obtained analogously to \eq{FKS} and \eq{I} by choosing $ \A = \sqrt{i} 
\, (\p - a(\q))$ and
$\B = i v( \q)$ in \eq{FK}. Its validity obviously requires some analyticity 
assumptions on the
potentials and the (initial) wavefunction. For rigorous discussions of 
\eq{uni} (in the case $a=0$), see \cite{Ha94, AzDo85} and 
the book
\cite{HiKu93} on a related theme.
\ei\pagebreak

\subsection{On the validity of the standard Feynman--Kac formula}

Equations \eq{I} and \eq{II} are the most popular versions of the 
\FK\ formula. It is therefore
fortunate that there is a wealth of information about the conditions 
under which they are rigorously
valid. We take the opportunity to recall some rather weak assumptions 
on the potentials $a$ and $v$ which are
sufficient \cite{Sim79, Sim82,AiSi82, CyFrKiSi87, BrHuLe}. To this end, 
let $L_{\mbox{\scriptsize\it loc}}^p(\rz^d)$, $p \in [1,\infty[$, denote 
the space of
complex--valued functions $\varphi: \rz^d \rta \cz$ with $\int_\Lambda 
\! dq\, \abs{\varphi(q)}^p <
\infty $ for all compact $\Lambda \subset \rz^d$. Let $v_\pm(q) := 
\sup \{0, \pm v(q)\} $ be the positive
and negative part of the scalar potential $v = v_+ - v_-$. 

In a first step we show the existence of the Wiener integral in
\eq{I} for $a = 0$ and $v_-$ belonging to the Kat\^o class 
${\cal K}(\rz^d)$, that is,
$\lim_{t\downarrow 0} \kappa_t(v_-) = 0$ where
\beq
   \kappa_t(u):= \sup_{x \in \krz^d} \IDw \int_0^t \! ds \, \abs{u(x + w(s))} 
        = \sup_{x \in \krz^d} \int_0^t \! ds \, \left(
\ep{-s \p^2 /2} \abs{u}\right) (x) \, .
\eeq
Due to the Markov property of $\Dw$ one may assume without loss of 
generality $t > 0$ sufficiently
small, so that $\kappa_t (v_-) < 1/2$. Then the following estimate holds
\beq \label{star}
\absq{ \IDw \exp\left\{ - \Ids{t} v(q + w(s)) \right\} \psi(q + w(t)) }  
              \le \frac{ \int dx \abs{\psi(x)}^2}{\DS (2 \pi t)^{d/2} 
(1 - 2\kappa_t(v_-))} \, .
\eeq
Its proof starts from employing the Cauchy--Schwarz--Bunyakovski  inequality
\beq \label{CS}
 \absq{\langle M_t(q,v) \psi(q + w(t))  \rangle }
  \le \langle M_t(q,2v) \rangle
       \langle \abs{\psi(q + w(t))}^2 \rangle 
\eeq
with respect to $\Dw$, where we have introduced the (multiplicative) functional
\beq
M_t(q,v) := \exp\left\{-\Ids{t} v(q + w(s))\right\} \, .
\eeq
While the second factor on the right--hand side of \eq{CS} is easily 
seen not to exceed $ \int dx \abs{\psi(x)}^2 / (2 \pi t)^{d/2}
< \infty$, the first factor is estimated using $ - v \le v_-$ and 
{\it Khas'minskii's Lemma} \cite{Kas59, AiSi82, Sim82}
  $   \langle M_t(q,-v_{-}) \rangle \le (1 - \kappa_t(v_-))^{-1} $.

Now turning on the vector potential $a$, the conditions 
$a^2,\nabla \! \cdot \! a \in L_{\mbox{\scriptsize \it loc}}^1(\rz^d)$ 
are sufficient for the existence of the Stratonovich
stochastic integral in \eq{I} (and hence of the corresponding 
$\a$-stochastic integral for arbitrary
$\a$). Since it appears there as the phase of a complex number 
of absolute value one, the Wiener
integral in \eq{I} remains bounded from above by the right--hand 
side of \eq{star}. In other words, the 
diamagnetic inequality \cite{Sim79} has extended \eq{star} to non--vanishing 
$a$. 

In order to guarantee equality \eq{I}, at least for Lebesgue--every $q$, 
one needs a condition on $v_+$, in addition to the above assumptions 
on $a$ and $v_-$.
For example, if $v_+ \in L_{\mbox{\scriptsize \it loc}}^1(\rz^d)$ the 
Hamiltonian $\H$ can be precisely
defined in the sense of quadratic forms as a self--adjoint operator  on 
$L^2(\rz^d)$ bounded from below and \eq{I}
holds.
Under the stronger assumptions $v_- \in {\cal K}(\rz^d) 
\cap L_{\mbox{\scriptsize \it loc}}^2(\rz^d)$ and
 $v_+, a^2, \nabla \! \cdot \! a \in L_{\mbox{\scriptsize \it loc}}^2(\rz^d)$
the space of arbitrarily often
differentiable complex--valued functions with compact support, 
$C^\infty_0(\rz^d)$,
can even be identified as an operator domain of essential self--adjointness.
\pagebreak

Pointwise validity of \eq{I} is assured by the assumptions 
$v_- \in {\cal K}(\rz^d)$ and $v_+$, $a^2$, 
$\nabla \! \cdot \! a \in {\cal K}_{\mbox{\scriptsize \it loc}}(\rz^d) \subset
L_{\mbox{\scriptsize \it loc}}^1(\rz^d)$, where by definition $u \in 
{\cal K}_{\mbox{\scriptsize \it loc}}(\rz^d) $ if 
$ u \varphi \in {\cal K}(\rz^d) $ for all $\varphi \in C_0^\infty(\rz^d)$. 
Under these assumptions the semigroup $\exp(-t\H)$ maps $L^2(\rz^d)$ into 
its subspace of functions which are
continuous (and bounded). As a consequence, both sides in \eq{I} are 
continuous and hence coincide for
all $q$. Even more, the semigroup possesses then for each $t>0$ an integral 
kernel 
$(q,q') \mapsto \bra{q} \exp(-t\H) \ket{q'}$ which is jointly continuous in 
$(t,q,q')$ for $t>0$ and $q,q' \in \rz^d$. It can be expressed by the 
right--hand side of \eq{II},
provided the delta function is understood in the sense of Donsker or Hida 
(see e.g.\ \cite{HiKu93}), or the integration
$\IDw \delta{(w(t)+q-q')(\cdot)}$ is interpreted as a short--hand for 
averaging with respect to the Brownian
bridge \cite{Sim79, Roe94, BrHuLe} from $(0,q)$ to $(t,q')$.

An  important example of a scalar potential, for which \eq{I} holds 
pointwise and the right--hand side of \eq{II}
defines the corresponding continuous integral kernel, is the  
inverse--distance potential $v(q) = - \gamma / \abs{q}$, $\gamma \in \rz$,
for $d \ge 2$. Complications in the case $d = 1$ are treated in 
\cite{FiLeMu95}.


\section{Phase--space path integrals for non--standard\protect\newline 
Hamil\-to\-nians}

Phase--space path integrals serve to provide a generalization of the 
standard \FK\ formula to
non--standard Hamiltonians on $L^2(\rz^d)$, that is, Hamiltonians  more 
general than
 $ \H  =  \half \left(\p - a(\q)\right)^2 + v(\q) $.
This idea dates back to Feynman \cite{Fey51} and is still used in quantum 
physics (see, e. g. \cite[Ch. 9]{Wei96}).
Any attempt to put it on safer grounds has to face the problem that there is 
no unique 
correspondence between Hilbert--space operators and phase--space functions.
Until now the most popular method to give meaning to Feynman's heuristic idea 
of phase--space
path integrals employs lattice or time--slicing prescriptions. Since the 
seventies it is well-known
\cite{Ali72, Dow76, LeSc77, Ber80, LaRoTi82, Pro82} that this method 
reflects the above non--uniqueness by a multiplicity of prescriptions, 
each linked to a 
particular way of choosing the
order for the factors in a product of non--commuting operators. In our 
opinion the most systematic way
to keep track of this is to proceed via linear phase--space representations 
(also known as
symbol--calculus in the theory of pseudo--differential operators) and a 
generalized Lie--Trotter
formula. In this section we are going to briefly summarize such a
procedure, restricting ourselves for clarity to certain one--parameter 
subclasses among phase--space representations and corresponding lattice 
prescriptions.

Following \cite{LeSc77} we associate with a given Hamilton operator $\H$ on 
$L^2(\rz^d)$ a
one--parameter family of {\em symbols}, that is, functions on classical phase 
space $\rz^d \times \rz^d$
defined by
\vspace*{-1.5mm}
\beq
     H_{\a}(p,q):=\int \!dx\, \ep{i p \! \cdot \! x} \bra{q-(1-\a)x} \H 
\ket{q+\a x} \, ,   \quad  \a \in [0,1]\, .
\eeq
For each fixed $\a$ the linear mapping  $\H \mapsto H_\a$ 
of Hilbert--space operators to phase--
\newpage
\noindent
space functions
can be inverted according to $\left\{H_\a(\p,\q)\right\}_\a = \H$. Here the 
mapping
\beq
     H \mapsto \left\{H(\p,\q)\right\}_\a 
            := \int \frac{dp dq}{(2 \pi)^d} \ep{i (\P  - p) \! \cdot \! q} 
H(p, \Q + \a q) \, .
\eeq
is a quantization or {\it operator--ordering scheme} \cite{AgWo70}
for phase--space functions $H$
which gives for $\a = 0, 1/2, 1$ anti--standard ($\p$ left of $\q$), 
Weyl--Wigner
(totally symmetrized) and standard ($\p$ right of $\q$) ordering, respectively.
A nice illustration is already provided by the example of a  standard 
Hamiltonian
\beqa \label{std}
   \H & = & \half \left(\p - a(\q)\right)^2 + v(\q) \\
   H_\a(p,q) & = & \half \left(p - a(q)\right)^2 + i \left(\a - \half\right) 
(\nabla \cdot a)(q) + v(q) \, . \vspace*{-1mm}
\eeqa
\mbox{}\vspace*{-2mm}%
Note that the {\em $\a$--symbol} $H_\a$ is real for all Hermitian $\H$ if 
and only if $\a = 1/2 $.
\vspace*{1mm}

We associate with a given $\H$ for each $\a$ the
operator--valued function
\beq
\R_{\a}(t) := \int \frac{dp dq}{(2 \pi)^d} \ep{i (\P - p) \! \cdot \! q} 
\exp{\left\{ -t  H_{\a}(p, \Q + \a q )\right\}}
\, , \qquad \, t \ge 0\, .
\eeq
It provides a short--time approximation to the semigroup 
$\exp(-t\H)$ in the sense that
\vspace*{-0.5mm}
\beq
\R_{\a}(0) = \one     \, , \qquad      \dot\R_{\a}(0) := \left.
\mbox{$\dby{t}\R_\a(t)$}\right|_{t=0}  = - \H \, .
\eeq
The idea now is to represent $\exp(-t\H)$ for arbitrary $t \ge 0$ as 
$\left[ \R_{\a}\!\left(\frac{t}{n}\right)\right]^n $ in the 
limit\linebreak\vspace*{1mm}%
$ n \rta \infty$ by writing
\beqa
\exp\left\{-t\H\right\}  & = & \exp\Bigl\{t \dot\R_{\a}(0)\Bigr\} = 
                    \lim_{n \rightarrow \infty} 
\left[\R_{\a}\!\left(\mbox{$\frac{t}{n}$}\right)\right]^n \\
     &  = & \lim_{n \rightarrow \infty} \int \frac{dp^{(n)} dq^{(n)}}{
     (2 \pi)^d} \cdot \dots \cdot \int \frac{dp^{(1)} dq^{(1)}}{(2 \pi)^d}  
            \ep{-i q^{(n)} \! \cdot \! \P} \label{disc}\\
            &   & \qquad   \times \exp{\Bigl\{ \sum_{\n=1}^{n} \Bigl[ i 
p^{(\n)} \! \cdot \! (q^{(\n)} -  q^{(\n-1)}) - \frac{t}{n} 
H_{\a}(p^{(\n)}, \Q + q^{(\n,\a)}) \Bigr] \Bigr\} }  \nonumber
\eeqa
where $q^{(0)} := 0$ and the so--called {\em $\a$--point} between 
$q^{(\n)}$ and $q^{(\n-1)}$ is defined as 
$ q^{(\n,\a)} := \a q^{(\n)} + (1-\a) q^{(\n-1)} $. It is customary and mnemonically convenient to use a continuum notation for the right--hand side so that
\vspace*{0mm}
\eqbox{ \label{cont} \mbox{\rule[-.9em]{0em}{2.5em}}
\ep{-t\H} = \int_{\a}\!{\cal{D}}\eta {\cal{D}}\xi \, \ep{-i \xi(t) 
\! \cdot \! \P} \exp{\left\{\int_0^t \! ds \Bigl[ i \eta(s) \cdot 
\dot\xi(s) -  
H_{\a}(\eta(s), \Q + \xi(s)) \Bigr] \right\}} \, .
} 
This expression is interpreted as a formal integral over all paths $s 
\mapsto (\eta(s), \xi(s))$ in phase--space $\rz^d \times \rz^d$ with 
$\xi(0)=0$. The 
subscript $\a$ attached to the integration sign reminds that it is essential 
to strictly use the {\it $\a$--point discretization} together with the 
$\a$--symbol \cite{LeSc77, LaRoTi82, Pro82}.
Otherwise there is no guarantee that the phase--space path integral 
represents the given semigroup. Conversely, replacing $H_{\a}$ in 
\eq{cont} by a given 
classical 
Hamilton function $H$ leads to the semigroup generated by the $\a$--ordered 
Hamilton operator $\{H(\p,\q)\}_{\a}$.
In case of the standard Hamiltonian \eq{std} the momentum variables in 
\eq{disc} can be integrated out yielding a discretized expression for the 
\FK--$\a$ 
formula (cf.\
\eq{FKS} and \eq{conversion}). Thereby one recognizes the well--known fact
that the $\a$--point discretization underlies the definition of the 
$\a$--stochastic
integral \cite{McG74} \cite[p.\ 136]{ReYo94}, which hence is in one--to--one 
correspondence 
with the $\a$--{\it ordering scheme} (see also \cite{LaRoTi82}). This is 
reflected by the conversion formula
\vspace*{-0.8mm}
\beq
   \half\left(\p g(\q) + g(\q) \p \right) = \left\{ \p g(\q)\right\}_\a + i 
\left(\half - \a \right) \left( \nabla \! \cdot \! g \right)(\q) \\[-2mm]
\eeq\mbox{}\vspace*{-2mm}%
similar to \eq{conversion}.\vspace{1mm}

The basic ingredient for the validity of \eq{disc} is the {\em generalized 
Lie--Trotter formula}
\vspace*{-0.5mm}
\beq \lim_{n\rta\infty} \left[\F\left(\mbox{$\frac{t}{n}$}\right)\right]^n = 
\exp\Bigl\{ t \dot \F(0) \Bigr\} \qquad \mbox{if} \qquad  \F(0)  = \one \, . 
\eeq
It can easily 
be verified by a Taylor expansion if $\F(t)$ acts on a finite--dimensional 
Hilbert space. Suitable technical assumptions for the extension to 
the infinite--dimensional case are specified in \cite{Che68}, see also 
\cite{Exn85}. The 
(semigroup version of the)
original Lie--Trotter formula (see, e.g.\ \cite{Sim79}) corresponds to
the choice $\F(t) = {\rm e}^{-t\A} {\rm e}^{-t\B}$ for operators $\A$ and 
$\B$ obeying suitable conditions. Nelson \cite{Nel64} was the first to use it 
explicitly in the context of
path integration. Note that without the presence of the Wiener measure the 
application of the original Lie--Trotter formula
with $\A = \p^2/2$ and $\B = v(\q)$ is of much less analytical value than 
the standard Feynman--Kac formula (for $a = 0$). In fact, one
can use the latter to prove norm--convergence in the former \cite{IchI}.

An interesting question is whether it is possible to construct a measure on  
a suitable set of paths in phase space in order to
benefit from general integration theory also in the case of non--standard 
Hamiltonians. One possibility is offered by Wiener--regularized expressions
in the spirit of \cite{DaKl85}. Another one derives from \eq{product} for 
Hamiltonians of the form
$\H =  h_+(\p) h_-(\q) + h_-(\q) h_+(\p)$. Both methods may also be used for 
systems with other than canonical degrees of freedom,
e.g.\ spin systems.

{\ixpt
We want to thank the local organizing committee for making possible this 
stimulating conference.\\[-4pt]
H. L. acknowledges support by the Heisenberg--Landau programme. B. B. and 
S. W. want to thank the\\[-4pt] 
Studienstiftung des deutschen Volkes for support.}


\end{document}
